\documentclass[11pt]{article}

\usepackage{amsmath}

\newcommand{\N}{N\raise.7ex\hbox{\underline{$\circ $}}$\;$}

\begin{document}

\begin{center}

{\bf V.M. Red'kov \\
Dirac-K\"{a}hler field, spinor technique, \\
and 2-potential approach
to electrodynamics with two charges }\\
B.I. Stepanov Institute of Physics \\
National Academy of Sciences of Belarus  \\
68 Nezavisimosti av., Minsk, 220072 BELARUS \\
redkov@dragon.bas-net.by

\end{center}

\begin{quotation}
From the 16-component Dirac-K\"{a}hler field theory, spinor
equations  for two types of massless  vector  photon fields with
different parities have been  derived. Their equivalent tensor
equations in terms of the strength tensor $F_{ab}$ and respective
4-vector  $A_{b}$ and 4-pseudovector  $\tilde{A}_{b}$ depending on
intrinsic photon parity are derived; they include additional
sources, electric 4-vector $j_{b}$ and magnetic 4-pseudovector
$\tilde{j}_{b}$. The theories of two types of photon fields are
explicitly uncoupled, their linear combination through summing or
subtracting results in Maxwell electrodynamics  with electric and
magnetic charges in 2-potential approach. So the problem of
existence of magnetic charge can be understood as a super
selection rule for different photon fields in  intrinsic parity.
The whole analysis is extended  straightforwardly to a curved
space-time background. In the frames of that extended Maxwell
theory, the known electromagnetic duality is  described as a
linear transformation mixing the field variables referred  to
photons with different parities. That extended dual transformation
concerns both strength tensors and 4-potentials  $A_{b},
\tilde{A}_{b}$.

\end{quotation}

Ppacs:  02.20, 03.50.De, 03.65.Fd, 03.65.Pm


\section{Introduction}

\hspace{5mm} The bibliography on Maxwell electrodynamics is
enormous, the subject attracts attention  up to the present time:

\vspace{4mm}

\noindent Lorentz  \cite{1886-Lorentz}, Poincar\'e
\cite{1905-Poincare}, Einstein \cite{1905-Einstein}, Silberstein
\cite{1907-Silberstein},  Minkowski \cite{1908-Minkowski},
Abraham \cite{1909-Abraham}, Bateman \cite{1909-Bateman}, Cunningham
\cite{1910-Cunningham},   Lanczos \cite{1919-Lanczos}, Lewis \cite{1910-Lewis},
Marcolongo  \cite{1912-Marcolongo},
Gordon  \cite{1923-Gordon}, Tamm -- Mandelstam \cite{1924-Tamm}
Rainich \cite{1925-Rainich}, Tamm  \cite{1929-Tamm}, Uhlenbeck --  Laporte
\cite{1931-Uhlenbeck-Laporte}, Juvet  \cite{1932-Juvet-Schidlof}, Abraham -- Becker
\cite{1932-Abraham-Becker}-\cite{1933-Abraham-Becker},
Frenkel \cite{1935-Frenkel}, Mercier
\cite{1935-Mercier}-\cite{1949-Mercier}, Rumer   \cite{1936-Rumer},
Stratton  \cite{1941-Stratton}, Rozen \cite{1952-Rosen},
Gursey  \cite{1954-Gursey},   Gupta \cite{1954-Gupta},
Lichnerowicz  \cite{1955-Lichnerowicz}, Novacu \cite{1955-Novacu}
 Borgardt \cite{1957-Borgardt}-\cite{1958-Borgardt},
Moses \cite{1958-Moses}-\cite{1959-Moses},
Panofsky -- Phillips  \cite{1962-Panofsky-Phillips},
  Post \cite{1962-Post},
Rosen \cite{1962-Rosen}-\cite{1973-Joe-Rosen},
Lipkin  \cite{1964-Lipkin},
Ellis  \cite{1964-Ellis}, Penney \cite{1964-Penney}, Zeldovich \cite{1965-Zeldovich},
Candlin  \cite{1966-Candlin},
Strazhev -- Tomil'chik
\cite{1968-Strazhev}-\cite{1971-Strazhev}-\cite{1972-Strazhev(1)}-\cite{1972-Strazhev(2)}-\cite{1974-Strazhev}-\cite{1975-Strazhev-Tom},
Exton -- Newman -- Penrose \cite{1969-Exton-Newman-Penrose},
Carmeli \cite{1969-Carmeli}, Pestov   \cite{1971-Pestov}, Landau-Lifshitz
\cite{1973-Landau},  Weingarten  \cite{1973-Weingarten}, Newman
\cite{1973-Newman},   Mignani -- Recami -- Baldo  \cite{1974-Mignani-Recami-Baldo},
Jackson  \cite{1974-Jackson},  Fushchich -- Nikitin
\cite{1974-Fushchich(1)}-\cite{1983-Fushchich(2)},
 Frankel  \cite{1974-Frankel},
Edmonds  \cite{1975-Edmonds},  Silveira \cite{1980-Silveira},
Venuri \cite{1981-Venuri},
Chow \cite{1981-Chow},
Cook \cite{1982-Cook(1)}-\cite{1982-Cook(1)},
Giannetto \cite{1985-Giannetto},
Y\'epez -- Brito -- Vargas \cite{1988-Nunez},
Kidd -- Ardini -- Anton \cite{1989-Kidd},
Recami \cite{1990-Recami},
 Krivsky -- Simulik  \cite{1992-Krivsky},
Inagaki \cite{1994-Inagaki},
Bialynicki-Birula \cite{1994-Bialynicki-Birula}-\cite{1996-Bialynicki-Birula}-\cite{2005-Birula},
Sipe \cite{1995-Sipe},
Ghose  \cite{1996-Ghose},
Gersten \cite{1998-Gersten},
 Esposito \cite{1998-Esposito},
Dvoeglazov  \cite{1998-Dvoeglazov} -\cite{2001-Dvoeglazov},
Kruglov \cite{2002-Kruglov},
Gsponer \cite{2002-Gsponer},
Kravchenko \cite{2002-Kravchenko},
Varlamov \cite{2002-Varlamov},
Ivezi\'c \cite{2001-Ivezic}-\cite{2002-Ivezic(1)}-\cite{2002-Ivezic(2)}-\cite{2003-Ivezic}-\cite{2005-Ivezic(1)}-\cite{2005-Ivezic(2)}-\cite{2005-Ivezic(3)}-\cite{2006-Ivezic},
Donev -- Tashkova \cite{2004-Donev(1)}-\cite{2004-Donev(2)}-\cite{2004-Donev(3)},  Armour \cite{2004-Armour}.

\vspace{4mm}

In the context of studying the Maxwell equations as equations
involving a massless field with spin one and definite intrinsic
parity, the well-known field of Dirac-K\"{a}hler  (see the recent book \cite{2007-Strazhev} on the subject and references therein)
 is of primary importance.
This  field obeys  a simple wave equation in spinor basis, this
field reduces to more simple theories for scalar or vector
particles with different intrinsic parities when  adding  linear
constraints on  field variables
 ( see in  \cite{1989-Red'kov}-\cite{2000-Red'kov}-\cite{2005-Tokarevskaya-Red'kov}). The main goal of the present
paper is to investigate that  possibility in  the context of Maxwell theory.

\section{Dirac-K\"{a}hler composite boson, wave equation}

In the Minkowski space, the Diac-K\"{a}hler particle is described
by 16-component wave function, a 2-rang bispinor, or equivalent
set of tensor fields:
\begin{eqnarray}
 U(x) \; ,  \qquad
\mbox{or} \qquad \{\; \Phi (x), \Phi _{i}(x), \tilde{\Phi }(x),
     \tilde{\Phi }_{i}(x), \Phi _{mn}(x) \; \} \; ;
\nonumber
\end{eqnarray}

\noindent $\Phi (x)$ is a scalar, $\Phi _{i}(x)$ is a vector,
$\tilde{\Phi }(x)$ represents a pseudoscalar, $\tilde{\Phi
}_{i}(x)$ represents a pseudovector, $\Psi _{mn}(x)$ is an
antisymmetric tensor. To specify connection between  2-rank
bispinor $U(x)$ and the tensorial set, let us introduce
parametrization of $U(x)$  according to
\begin{eqnarray}
U(x) = \left [\;  - i \; \Phi (x) + \gamma^{l} \; \Phi _{l}(x) +
           i \; \sigma^{mn}\;  \Phi _{mn}(x) +  \gamma ^{5} \; \tilde{\Phi }(x) +
           i \; \gamma ^{l} \gamma ^{5} \; \tilde {\Phi }_{l}(x) \; \right ]\; E^{-1} \; ,
\nonumber
\\
\gamma^{5} = - i \gamma^{0} \gamma^{1} \gamma^{2} \gamma^{3} \; ,
\qquad
 \sigma^{ab} = {1 \over 4} (\gamma^{a}  \gamma^{b} - \gamma^{b} \gamma^{a}) \;  ;
\label{5.1}
\end{eqnarray}

\noindent where  $E$ stands for a metrical matrix in 4-spinor space:
\begin{eqnarray}
E =  \left | \begin{array}{cc}
               \epsilon   &   0  \\ 0   &   \dot{\epsilon}^{-1}
\end{array}  \right | =
     \left | \begin{array}{cc}
               \epsilon_{\alpha \beta}  &   0  \\
                0      &      \epsilon ^{\dot{\alpha}\dot{\beta}}
     \end{array}  \right | =
     \left | \begin{array}{cc}
              i  \sigma^{2}  &   0  \\
                0      &    - i   \sigma^{2}
     \end{array}  \right |  ,
\nonumber
\\
E^{-1} =  \left | \begin{array}{cc}
               \epsilon^{-1}   &   0  \\ 0   &   \dot{\epsilon}
\end{array}  \right | =
     \left | \begin{array}{cc}
               \epsilon^{\alpha \beta}  &   0  \\
                0      &      \epsilon _{\dot{\alpha}\dot{\beta}}
     \end{array}  \right | =
     \left | \begin{array}{cc}
              -i  \sigma^{2}  &   0  \\
                0      &    + i   \sigma^{2}
     \end{array}  \right | \; ,
\nonumber
\\
E^{2} = - I \; , \qquad \tilde{E} = - E \; , \qquad  \mbox{Sp} \;
E = 0 \; , \qquad
 \tilde{\sigma }^{ab} \;  E = - E \; \sigma ^{ab} \;  .
\label{5.2}
\end{eqnarray}

\noindent Inverse to  (\ref{5.1})  has the form
\begin{eqnarray}
\Phi _{l}(x) = {1\over 4} \; \mbox{Sp}  \; [ E \gamma _{l} U(x)]
\; , \qquad \tilde{\Phi }_{l}(x) = {1\over 4i}  \; \mbox{Sp} \; [E
\gamma ^{5}\gamma _{l} U(x)] \; ,
\nonumber
\\
\Phi (x) = {i \over 4} \; \mbox{Sp} \; [ E U(x) ] \; , \qquad
\tilde{\Phi }(x) = {1\over 4} \;  \mbox{Sp}\;  [E \gamma ^{5}
U(x)] \; ,
\nonumber
\\
\Phi _{mn}(x)= -{1 \over 2i} \; \mbox{Sp} \;  [E \sigma _{mn} U(x)
]\; .
\label{5.3}
\end{eqnarray}

Below we will use the  Weyl's basis:
\begin{eqnarray}
U  = \left | \begin{array}{cc}
\xi ^{\alpha \beta } & \Delta ^{\alpha}_{\;\;\; \dot{\beta }} \\
H_{\dot{\alpha}} ^{\;\;\;\beta}  & \eta
_{\dot{\alpha}\dot{\beta }}
\end{array} \right | \; , \qquad
\gamma^{a} = \left | \begin{array}{cc}
0 & \bar{\sigma}^{a} \\
\sigma^{a} & 0   \end{array} \right | \; , \qquad  \sigma^{a} = (I, \sigma^{k})\;
, \;  \bar{\sigma}^{a} = (I, -\sigma^{k})\; ,
\nonumber
\\
\sigma^{ab}  = {1 \over 4}  \; \left | \begin{array}{cc}
\bar{\sigma}^{a} \sigma^{b} - \bar{\sigma}^{b} \sigma^{a} & 0 \\
0  & \sigma^{a} \bar{\sigma}^{b} - \sigma^{b} \bar{\sigma}^{a}
\end{array} \right | =
\left | \begin{array}{cc} \Sigma^{ab} & 0 \\ 0  &
\bar{\Sigma}^{ab}  \end{array} \right | \; , \;
\gamma^{5} =
\left | \begin{array}{cc} - I & 0 \\ 0 & +I \end{array} \right | ,
\label{5.4}
\end{eqnarray}

\noindent in which decomposition  (\ref{5.1}) looks
\begin{eqnarray}
 \left | \begin{array}{cc}
\xi ^{\alpha \beta } & \Delta ^{\alpha}_{\;\;\; \dot{\beta }} \\
H_{\dot{\alpha}} ^{\;\;\;\beta}  & \eta _{\dot{\alpha}\dot{\beta
}}
\end{array} \right | =
\left | \begin{array}{cc} [\; (-i\; \Phi - \tilde{\Phi}) + i\;
\Sigma^{mn} \;\Phi_{mn} \; ] \; \epsilon^{-1}  &
\bar{\sigma}^{l} \; (\Phi _{l} + i\; \tilde{\Phi}_{l}) \; \dot{\epsilon} \\[2mm]
\sigma^{l}\;  (\Phi _{l} - i \; \tilde{\Phi}_{l}) \; \epsilon^{-1}
& [\; (-i\; \Phi + \tilde{\Phi}) + i \; \bar{\Sigma}^{mn} \;
\Phi_{mn} \;]\;  \dot{\epsilon}
\end{array} \right |.
\label{5.5}
\end{eqnarray}

Thus, 2-spinor and tensor descriptions of  the Dirac-K\"{a}hler field are determined by
\begin{eqnarray}
\Delta  = ( \;\Phi _{l}  + i \; \tilde{\Phi}_{l}  \; )
\sigma^{l}\dot{\epsilon } \; ,
\nonumber
\\
H = ( \;\Phi _{l} - i \; \tilde{\Phi }_{l} ) \; \bar{\sigma}^{l}
\epsilon ^{-1} \; ,
\nonumber
\\
\xi  = ( \; - i \; \Phi  - \tilde{\Phi}  + i \Sigma ^{mn} \; \Phi
_{mn}\;  ) \; \epsilon ^{-1} \; ,
\nonumber
\\
\eta  = ( \; - i \; \Phi  + \tilde{\Phi}  +
 i \bar{\Sigma}^{mn}\; \Phi _{mn}\; )\; \dot{\epsilon } \;  ;
\label{5.9a}
\end{eqnarray}

\noindent
and inverse relations are
\begin{eqnarray}
\Phi _{l} + i \; \tilde{\Psi} _{l}  = {1\over 2} \; sp\;  (\;
\dot{\epsilon }^{-1} \; \sigma _{l} \Delta  \; ) \; , \qquad \Phi
_{l} - i\; \tilde{\Psi }_{l}  =
 {1\over 2} \; sp \;  (\;  \epsilon \bar{\sigma}_{l} \; H \; ) \; ,
\nonumber
\\
-i \; \Phi  - \tilde{\Psi}   = {1\over 2} \; sp\;  (\;  \epsilon
\; \xi \;  ) \; , \qquad -i \;\Phi  + \tilde{\Psi }  = {1\over 2}
\; sp \; ( \; \dot{\epsilon }^{-1} \; \xi \; ) \; ,
\nonumber
\\
  - i \; \Phi ^{kl} + {1\over 2}\; \epsilon ^{klmn} \;\Psi _{mn} =
sp\; ( \; \epsilon \;  \Sigma ^{kl} \xi\; ) \; ,
\nonumber
\\
 - i\;  \Phi ^{kl} - {1\over 2}\; \epsilon ^{klmn}\; \Psi _{mn} =
sp \; ( \; \dot{\epsilon }^{-1} \; \bar{\Sigma}^{kl} \;  \xi \; )
\; .
\label{5.9b}
\end{eqnarray}

Now let us turn to the Dirac-K\"{a}hler wave  equation. In spinor basis it is
a Dirac-like equation for 2-rank bispinor:
\begin{eqnarray}
(i \gamma^{a} {\partial \over \partial x^{a}} - m) \; U(x) = 0 \;
;
\label{5.10a}
\end{eqnarray}

\noindent    or in 2-spinor form
\begin{eqnarray}
(A)\qquad i\sigma ^{a} \;  \partial _{a} \; \xi (x) = m \;   H(x)
\; , \qquad (A') \qquad i\bar{\sigma}^{a} \; \partial_{a} \;  H(x)
= m \; \xi (x) \; ,
\nonumber
\\
(B)\qquad i \bar{\sigma}^{a} \;  \partial_{a} \; \eta (x) = m \;
\Delta (x) \; , \qquad (B') \qquad i \sigma ^{a} \; \partial_{a}
\; \Delta (x)    = m \; \eta (x) \; .
\label{5.10b}
\end{eqnarray}

Equations   (\ref{5.10a}) -(\ref{5.10b})  are invariant under the Lorentz group.
First, let us consider continuous transformations from  SL(2.C).
 the Dirac-K\"{a}hler field transforms according to
\begin{eqnarray}
U'(x') = [\; S(k ,k^{*}) \otimes  S(k,k^{*}) \; ] \; U(x) \; ;
\nonumber
\\
\xi ' (x') = B(k) \xi (x)\tilde{B}(k) \; , \qquad  \Delta'(x') =
B(k) \Delta (x) \tilde{B}(\bar{k}^{*}) \; ,
\nonumber
\\
H'(x') = B(\bar{k}^{*}) H (x) \tilde{B}(k)\; , \qquad \eta'(x') =
B(\bar{k}^{*}) \eta (x) \tilde{B}(\bar{k}^{*}) \; .
\label{5.11b}
\end{eqnarray}

\noindent We should find equation satisfied by primed quantities.
From eq.  (A) in (\ref{5.10b}) it follows
\begin{eqnarray}
(A) \qquad i\sigma ^{a} \; \partial_{a} \; B^{-1}(k) \xi'
(x')\tilde{B}^{-1}(k)
 = m \;  B(k^{*})  H'(x') \tilde{B}^{-1}(k)  \; , \qquad \Longrightarrow
\nonumber
\\
i \; B^{-1}(k^{*}) \sigma ^{a}  B^{-1}(k) \; \partial_{a} \; \xi'
(x')   = m \;   H'(x')
\nonumber
\end{eqnarray}

\noindent and further, taking into account identity
\begin{eqnarray}
B^{-1}(k^{*}) \sigma ^{a}  B^{-1}(k) = \sigma^{b}
L_{b}^{\;\;\;a}(k,k^{*}) \; ,
\label{5.12a}
\end{eqnarray}

\noindent we get
\begin{eqnarray}
(A) \qquad i\sigma ^{b} \;  \partial \;' _{b} \; \xi' (x') = m \;
H'(x') \; , \qquad \mbox{where } \qquad L_{b}^{\;\;\;a}(k,k^{*})
\partial_{a} = \partial \; '_{b} \; .
\label{5.12b}
\end{eqnarray}

\noindent Analogously, from eq. $(A')$ in  (5.10b) it follows
\begin{eqnarray}
(A') \qquad i \bar{\sigma} ^{a} \; \partial_{a} \; B(k^{*}) H'
(x')\tilde{B}^{-1}(k)
 = m \;  B^{-1}(k)  \xi'(x') \tilde{B}^{-1}(k)  \; , \qquad \Longrightarrow
\nonumber
\\
i \; B(k) \bar{\sigma} ^{a}  B(k^{*}) \; \partial_{a} \; H' (x')
\;
 = m \;   \xi'(x')\; ;
\nonumber
\end{eqnarray}

\noindent from whence with the use of identity
\begin{eqnarray}
B(k) \bar{\sigma} ^{a}  B(k^{*}) = \bar{\sigma}^{b}
L_{b}^{\;\;\;a}(k,k^{*})\; ,
\label{5.13a}
\end{eqnarray}

\noindent  we arrive at
\begin{eqnarray}
(A') \qquad i \bar{\sigma} ^{b} \;  \partial \;' _{b} \; H' (x') =
m \;  \xi'(x')  \; .
\label{5.13b}
\end{eqnarray}

\noindent
In the same manner one may prove invariance of remaining equations $(B)$ and $(B')$ in  (\ref{5.10b}).
Note that two identities  (\ref{5.12a}) and (\ref{5.13a}) in 2-form can be rewritten in 4-form:
\begin{eqnarray}
S(k,\bar{k}^{*})  \gamma^{a} S^{-1} (k,\bar{k}^{*})  = \gamma^{b}
\; L_{b}^{\;\;\;a}(k,k^{*})\; ,
\label{5.14}
\end{eqnarray}

\noindent with the help of which one can easily demonstrate
invariance of the Dirac-K\"{a}hler equation in 4-spinor form.
Invariance of the   Dirac-K\"{a}hler equation under two discrete
transformations
\begin{eqnarray}
M = i \gamma^{0} , \qquad \left | \begin{array}{cc}
\xi' & \Delta' \\
H'  & \eta'
\end{array} \right | =
(M \otimes M) \left | \begin{array}{cc}
\xi & \Delta \\
H  & \eta
\end{array} \right | =
\left | \begin{array}{cc}
-\eta & -H \\
-\Delta   & -\xi
\end{array} \right | \; ,
\nonumber
\\
N = \gamma^{0} \gamma^{5}  , \qquad \left | \begin{array}{cc}
\xi' & \Delta' \\
H'  & \eta'
\end{array} \right | =
(N \otimes N) \left | \begin{array}{cc}
\xi & \Delta \\
H  & \eta
\end{array} \right | =
\left | \begin{array}{cc}
\eta & -H \\
-\Delta   & \xi
\end{array} \right | \; ,
\label{5.15}
\end{eqnarray}

\noindent is readily proved with the use of two identities:
\begin{eqnarray}
M \gamma^{a} M^{-1} = \left | \begin{array}{cc} 0 & \sigma^{a}  \\
\bar{\sigma}^{a}  & 0 \end{array} \right | = \gamma^{b} \;
L_{b}^{\;\;(P)a} , \qquad N \gamma^{a} N^{-1} = \left |
\begin{array}{cc} 0 & -\sigma^{a}  \\ -\bar{\sigma}^{a}  & 0
\end{array} \right | = \gamma^{b} \; L_{b}^{\;\;(T)a} \;.
\label{5.16}
\end{eqnarray}

Now let us derive a set of tensor equations equivalent to the above spinor form.
To this end, the spinor  equation is to be presented  as
\begin{eqnarray}
 i \gamma ^{c} \;  \partial_{c} \;
(\; - i \Phi  +  \gamma ^{l} \; \Phi _{l} + i \sigma ^{mn} \; \Phi
_{mn} + \gamma ^{5} \tilde{\Phi} + i \gamma ^{l}\gamma ^{5}
\tilde{\Phi }_{l} )\; E^{-1} \; -
\nonumber
\\
- m \; (\; - i \Phi  + \gamma ^{l} \Phi _{l} + i \sigma ^{mn} \Phi
_{mn} +
 \gamma ^{5} \tilde{\Phi} + i \gamma ^{l} \gamma ^{5} \tilde{\Phi }_{l}\; )
 E^{-1}\;   = 0\; ,
\label{5.17}
\end{eqnarray}

\noindent from whence with the use of the formulas
\begin{eqnarray}
\mbox{Sp}\; \gamma^{a} =  0 \; , \qquad
\mbox{Sp}\; \gamma^{5} =  \mbox{Sp}\; \left | \begin{array}{cc}
-I &  0  \\
0 & +I
\end{array} \right | = 0 \; ,\qquad
\mbox{Sp}\; ( \gamma^{5} \gamma^{a} )= \mbox{Sp}\; \left |
\begin{array}{cc}
0 &  -\bar{\sigma}^{a} \\
+\sigma^{a} & 0
\end{array} \right | = 0 \; ;
\nonumber
\\
\mbox{Sp} \; (\sigma^{a} \bar{\sigma}^{b} ) =
\mbox{Sp} \; (\bar{\sigma}^{a} \sigma^{b} ) =   2 \; g^{ab}
\; , \qquad
\mbox{Sp} \; ( \gamma^{a} \gamma^{b} )  = \mbox{Sp}\; \left |
\begin{array}{cc}
\bar{\sigma}^{a} \sigma^{b} & 0 \\
0 & \sigma^{a} \bar{\sigma}^{b}
\end{array} \right | = 4 g^{ab} \; ,\;\;
\nonumber
\\
\mbox{Sp} \; ( \gamma^{5} \gamma^{a} \gamma^{b} ) =  0 \; , \qquad
\mbox{Sp}\; ( \gamma^{c} \gamma^{a} \gamma^{b} ) = \mbox{Sp}\;
\left | \begin{array}{cc}
0  & \bar{\sigma}^{c} \sigma^{a} \bar{\sigma}^{b} \\
\sigma^{c} \bar{\sigma}^{a} \sigma^{b}  & 0
\end{array} \right | = 0 \;  , \;
\mbox{Sp}\; (\gamma^{5} \gamma^{c} \gamma^{a} \gamma^{b} ) = 0 \;  ;
\nonumber
\\
\mbox{Sp} \; ( \bar{\sigma}^{d} \sigma^{c} \bar{\sigma}^{a}
\sigma^{b} ) = 2\; (\; g^{dc} g^{ab} - g^{da} g^{cb} + g^{db}
g^{ca} - i\; \epsilon^{dcab} \; )\; ,
\nonumber
\\
\mbox{Sp} \; (\sigma^{d} \bar{\sigma}^{c} \sigma^{a}
\bar{\sigma}^{b} ) = 2\; (\; g^{dc} g^{ab} - g^{da} g^{cb} +
g^{db} g^{ca} + i\; \epsilon^{dcab} \;)\; ,
\nonumber
\\
\mbox{Sp}\; ( \gamma^{d} \gamma^{c} \gamma^{a} \gamma^{b} ) =
\mbox{Sp}\; \left | \begin{array}{cc}
\bar{\sigma}^{d} \sigma^{c} \bar{\sigma}^{a} \sigma^{b}  & 0\\
0 &  \sigma^{d} \bar{\sigma}^{c} \sigma^{a} \bar{\sigma}^{b}
\end{array} \right | = 4\; (\; g^{dc} g^{ab} - g^{da} g^{cb} + g^{db} g^{ca} \;) \; ,
\nonumber
\\
\mbox{Sp}\; ( \gamma^{5} \gamma^{d} \gamma^{c} \gamma^{a}
\gamma^{b} ) = \mbox{Sp}\; \left | \begin{array}{cc}
- \bar{\sigma}^{d}  \sigma^{c}  \bar{\sigma}^{a}  \sigma^{b}  & 0 \\
0  &   \sigma^{d}  \bar{\sigma}^{c}  \sigma^{a}  \bar{\sigma}^{b}
\end{array} \right | =  4i\; \epsilon^{dcab}   \; .
\nonumber
\end{eqnarray}

\noindent  and also
\begin{eqnarray}
\sigma^{a} \bar{\sigma}^{b} \sigma^{c} = \sigma^{a} g^{bc} -
\sigma^{b} g^{ac} + \sigma^{c} g^{ab} + i\; \epsilon^{abcd} \;
\sigma_{d} \; ,
\nonumber
\\
\bar{\sigma}^{a} \sigma ^{b} \bar{\sigma}^{c} = \bar{\sigma}^{a}
g^{bc} - \bar{\sigma}^{b} g^{ac} + \bar{\sigma}^{c} g^{ab} - i\;
\epsilon^{abcd} \; \bar{\sigma}_{d} \; , \qquad
\gamma^{a} \gamma^{b} \gamma^{c} = \left | \begin{array}{cc}
0 & \bar{\sigma}^{a} \sigma ^{b} \bar{\sigma}^{c} \\
\sigma^{a} \bar{\sigma}^{b} \sigma^{c}  &  0
\end{array} \right |
\nonumber
\end{eqnarray}

\noindent we readily arrive at
\begin{eqnarray}
\partial_{l} \Phi ^{l} +
  m\; \Phi = 0 \; ,
\qquad
\partial_{l}
 \tilde{\Phi }^{l}  + m \tilde{\Phi}  = 0 \; ,
\nonumber
\\
\partial^{k} \Phi  +
 \partial_{l} \Phi^{kl}  - m \Phi ^{k} = 0 \; ,
\nonumber
\\
\partial^{k} \tilde{\Phi} -
{1 \over 2} \epsilon ^{kcmn} \;
\partial_{c} \Phi _{mn}
 - m \tilde{\Phi }^{k}  = 0  \; ,
\nonumber
\\
\partial^{d} \Phi ^{k} -
\partial^{k} \Phi ^{d}  +
\epsilon ^{dkcl} \;  \partial_{c}
 \tilde{\Phi }_{l}  - m \Phi ^{dk}  = 0 \; .
\label{5.24}
\end{eqnarray}

\noindent It should be stressed that the  present context, the completely antisymmetric
  object  Levi-Civite  $\epsilon ^{abcd}$ appears
as a definite and fixed symbol, $\epsilon ^{0123}=+1$  and so on, without any presumed properties  under  the Lorentz
transformations.

One may easy verify that in each equation in  (\ref{5.24})  are
linearly combined quantities of the  the same behavior under the
spinor covering  $G_{L}^{spin}$ of the full Lorentz group
$L_{+-}^{\uparrow \downarrow}$ ; at this  the  space-time
coordinates behave as follows
\begin{eqnarray}
(M) \qquad x^{M}_{l} = (+ \bar{\delta}_{l}^{\;\;k})\;  x_{k} \; ,
\qquad (N) \qquad x^{N}_{l} = (- \bar{\delta}_{l}^{\;\;k})\; x_{k}
\; .
\label{5.25}
\end{eqnarray}

\section{On equations for particles with different parities
}

Let us consider eqs.  (\ref{5.24}) at four different additional constraints:
\begin{eqnarray}
\underline{S=0} \qquad \qquad \tilde{\Phi } =0 \; , \qquad
\tilde{\Phi }_{\alpha }=0 \; , \qquad  \Phi _{\alpha \beta } = 0
\; ,
\nonumber
\\
\partial^{l} \Phi_{l} + m \Phi  = 0 \; , \qquad 0=0 \; ,
\nonumber
\\
\partial_{l} \Phi  - m \; \Phi _{l} = 0 \; , \qquad 0=0 \; ,
\nonumber
\\
\partial^{d} \Phi ^{k} -
\partial^{k} \Phi ^{d}  = 0 \;  ;
\label{6.2b}
\end{eqnarray}
\begin{eqnarray}
\underline{S = \tilde{0}} \qquad \qquad \Phi  =0 \; , \qquad  \Phi
_{\alpha }=0 \; , \qquad  \Phi _{\alpha \beta } = 0 \; ;
\nonumber
\\
0=0 \; , \qquad
\partial^{l} \tilde{\Phi }_{l} +
 m \; \tilde{\Phi } = 0 \; ,
\nonumber
\\
0=0\; , \qquad
\partial ^{k } \tilde{\Phi } - m \tilde{\Phi }^{k}  = 0 \; ,
\nonumber
\\
\epsilon^{dkcl} \partial_{c} \Phi_{l} = 0 \; ;
\label{6.3b}
\end{eqnarray}
\begin{eqnarray}
\partial^{l} \tilde{\Phi }_{l} +
 m \; \tilde{\Phi } = 0 \; ,  \qquad
\partial ^{k } \tilde{\Phi } - m \tilde{\Phi }^{k}  = 0 \; .
\nonumber
\\
\underline{S=1} \qquad \qquad \Phi =0 \;  , \qquad \tilde{\Phi}=0
\; , \qquad \tilde{\Phi}_{l}=0 \; ;
\label{6.4a}
\end{eqnarray}
\begin{eqnarray}
\underline{S=\tilde{1}:} \qquad \qquad \Phi =0 , \qquad
\tilde{\Phi}=0, \qquad \Phi_{l}=0 \; ,
\nonumber
\\
0=0\; , \qquad
\partial^{l} \tilde{\Phi}_{l} = 0 \; , \qquad
\partial^{l} \Phi_{kl} = 0 \; ,
\nonumber
\\
{1\over 2}\; \epsilon ^{kcmn}\;
\partial _{c} \Psi _{mn}  +
m \;\tilde{\Phi}^{k} = 0 \; , \qquad
 \epsilon ^{dkcl} \;
\partial_{c} \tilde{\Phi }_{l} - m \;\Phi ^{dk} = 0 \; .
\label{6.5b}
\end{eqnarray}

\noindent In the following  two scalar cases will be  omitted.
Let  us describe additional constraints  in spinor form
 The main relation referring  spinors to tensors is
\begin{eqnarray}
 \left | \begin{array}{cc}
\xi  & \Delta  \\
H & \eta  \end{array} \right | = \left | \begin{array}{cc} [\;
(-i\; \Phi - \tilde{\Phi}) + i\; \Sigma^{mn} \;\Phi_{mn} \; ] \;
(-i\sigma^{2})   &
\bar{\sigma}^{l} \; (\Phi _{l} + i\; \tilde{\Phi}_{l}) \; (+i\sigma^{2})  \\[2mm]
\sigma^{l}\;  (\Phi _{l} - i \; \tilde{\Phi}_{l}) \;
(-i\sigma^{2})   & [\; (-i\; \Phi + \tilde{\Phi}) + i \;
\bar{\Sigma}^{mn} \; \Phi_{mn} \;]\; (+i\sigma^{2})
\end{array} \right |
\label{6.7}
\end{eqnarray}

\noindent
For a vector particle, we will have ( the symbol of $tr$ stands for a matrix transposition)
\begin{eqnarray}
\underline{S = 1}  \qquad   \left | \begin{array}{cc}
\xi  & \Delta  \\
H & \eta  \end{array} \right | = \left | \begin{array}{cc}
 + \; \Sigma^{mn} \;  \sigma^{2} \;\Phi_{mn}    &
+i\; \bar{\sigma}^{l} \sigma^{2} \; \Phi _{l}    \\[2mm]
-i \; \sigma^{l} \sigma^{2}  \;  \Phi _{l} \;    & -\;
\bar{\Sigma}^{mn} \; \sigma^{2} \; \Phi_{mn}
\end{array} \right |\; ;
\nonumber
\\
\tilde{\Delta } = + H \; ,
 \; \tilde{\xi } = + \xi \; ,
 \; \tilde{\eta } = + \eta \; .
\label{6.10b}
\end{eqnarray}

\noindent
 A pseudovector case is given by
\begin{eqnarray}
\underline{S = \tilde{1}} \qquad  \qquad  \left | \begin{array}{cc}
\xi  & \Delta  \\
H & \eta  \end{array} \right | = \left | \begin{array}{cc}
 + \Sigma^{mn} \sigma^{2} \;\Phi_{mn} \;     &
-\; \bar{\sigma}^{l} \sigma^{2} \;  \tilde{\Phi}_{l}  \\[2mm]
-\; \sigma^{l}\sigma^{2} \;   \tilde{\Phi}_{l}   &
 -\; \bar{\Sigma}^{mn} \sigma^{2} \; \Phi_{mn}
\end{array} \right | \; ,
\nonumber
\\
 \tilde{\Delta } = - H \;
, \; \tilde{\xi } = + \xi\; ,   \; \tilde{\eta } = + \eta \; .
\label{6.12b}
\end{eqnarray}

\section{Massless vector particle, and Lorentz condition}

Bearing in mind a photon field, it is better to use the usual  notation:
\begin{eqnarray}
\underline{S=1} \qquad  \qquad \Phi_{l} \Longrightarrow A_{l} ,
\qquad \Phi_{kl} \Longrightarrow F_{kl} \; ,
\nonumber
\\
U(x) = \left [\;  + \gamma^{l} \; A_{l}(x) +
           i \; \sigma^{mn}\;  F _{mn}(x)\; \right ]\; E^{-1} \; ;
\nonumber
\end{eqnarray}

\noindent the wave equation in spinor form is
\begin{eqnarray}
(A)\qquad i\sigma ^{a} \;  \partial _{a} \; \xi  = m \;   H \; ,
\qquad (A') \qquad i\bar{\sigma}^{a} \; \partial_{a} \;  H = m \;
\xi  \; ,
\nonumber
\\
(B)\qquad i \bar{\sigma}^{a} \;  \partial_{a} \; \eta  = m \;
\Delta \; , \qquad (B') \qquad i \sigma ^{a} \; \partial_{a} \;
\Delta     = m \; \eta  \; .
\label{7.1a}
\end{eqnarray}

\noindent which is equivalent to the  tensor system:
\begin{eqnarray}
\partial^{l} A_{l} = 0 \; , \qquad 0=0\; , \qquad
\partial^{l} F_{k l } - m \; A_{k}   =  0 \;  ,
\nonumber
\\
\epsilon^{kcmn} \; \partial_{c} F_{mn} = 0 \; , \qquad
\partial_{d} A _{k} - \partial_{k} A_{d} -
 m F_{dk}  =0  \; .
\label{7.1b}
\end{eqnarray}

\noindent Tensors and spinors are referred by
\begin{eqnarray}
 \left | \begin{array}{cc}
\xi  & \Delta  \\
H & \eta  \end{array} \right | = \left | \begin{array}{cc}
 + \; \Sigma^{mn} \;  \sigma^{2} \;F_{mn}    &
+i\; \bar{\sigma}^{l} \sigma^{2} \; A_{l}    \\[2mm]
-i \; \sigma^{l} \sigma^{2}  \;  A_{l} \;    & -\;
\bar{\Sigma}^{mn} \; \sigma^{2} \; F_{mn}
\end{array} \right | \;  ,
\nonumber
\end{eqnarray}

\noindent or in a more detailed form
\begin{eqnarray}
\xi - \eta  = -2i \; (\; \sigma^{1} \;
F_{23} + \sigma^{2}\; F_{31} +
\sigma^{3} \; F_{12} \; ) \; \sigma^{2} \; ,
\nonumber
\\
\xi + \eta = 2\; ( \;\sigma^{1}\; F_{01} + \sigma^{2} \;F_{02} +
\sigma^{3} \; F_{03} \;)  \; \sigma^{2} \; ,
\nonumber
\\
\Delta =
 \left | \begin{array}{rr}
(A_{1} - i A_{2} )  &  (A_{0} - A_{3}) \\
-(A_{0} + A_{3})   & -(A_{1} + i A_{2})
\end{array} \right | ,
\qquad
H =  \left | \begin{array}{rr}
(A_{1} - i A_{2})  & -(A_{0} + A_{3}) \\
(A_{0} - A_{3} )  & -(A_{1} + i A_{2})
\end{array} \right | \;  .
\label{7.2}
\end{eqnarray}

Restriction to a massless vector particle is achieved as follows (see (\ref{7.1a}) and (\ref{7.1b}) ):
\begin{eqnarray}
\left. \begin{array}{ll}
(A) & \qquad i\sigma ^{a} \;  \partial _{a} \; \xi  = 0  \\[2mm]
(B) & \qquad i\bar{\sigma}^{a} \; \partial_{a} \; \eta  = 0
\end{array} \right \}
\qquad
 \Longleftrightarrow \qquad \;\;\;
\partial^{l} F_{kl} = 0 \; , \qquad
\epsilon^{kcmn} \partial_{c} F_{mn} = 0 \; ;
\label{7.5}
\\
\left. \begin{array}{ll}
(A') & \qquad i\bar{\sigma} ^{a} \;  \partial _{a} \; H  =  \;  \xi  \\[2mm]
(B') & \qquad i\sigma^{a} \; \partial_{a} \; \Delta  =   \; \eta
\end{array} \right \}
\qquad
 \Longleftrightarrow \qquad
 \partial^{l} A_{l} = 0 \; , \qquad \partial_{k} A_{l} - \partial_{l} A_{k} =  \; F_{kl} \; .
\label{7.6}
\end{eqnarray}

\noindent
One should note that spinor equations $(A'),(B')$ results in  the Lorentz condition.
With the use of 3-vector notation
\begin{eqnarray}
F_{01} = - E_{1} = +E^{1} \;, \qquad F_{02} = -E_{2} = + E^{2}\;,
\qquad F_{03} = -E_{3} = +E^{3} \;,
\nonumber
\\
F_{23} = B_{1}=B^{1} \;, \qquad F_{31} = B_{2}=B^{2} \;, \qquad F_{12} =
B_{3}=B^{3}\;,
\nonumber
\end{eqnarray}

\noindent eqs.  (\ref{7.5}) take the form
\begin{eqnarray}
\underline{(A),(B)} \qquad  \qquad
\partial_{1} E_{1} + \partial_{2} E_{2} + \partial_{3} E_{3}  = 0 \; ,\qquad
\partial_{1} B_{1} + \partial_{2} B_{2} + \partial_{3} B_{3}  = 0 \; ,
\nonumber
\\
\partial_{2} E_{3} - \partial_{3} E_{2} = -\partial_{0} B_{1} \; ,
\qquad
\partial_{3} E_{1} - \partial_{1} E_{3} = -\partial_{0} B_{2} \; ,
\qquad
\partial_{1} E_{2} - \partial_{2} E_{1} = - \partial_{0} B_{3} \; ,
\nonumber
\\
\partial_{2} B_{3} - \partial_{3} B_{2} = +\; \partial_{0} E_{1} \; ,
\qquad
\partial_{3} B_{1} - \partial_{1} B_{3} = +\; \partial_{0} E_{2} \; ,
\qquad
\partial_{1} B_{2} - \partial_{2} B_{1} = + \; \partial_{0} E_{3} \; ,
\nonumber
\end{eqnarray}

\noindent or in 3-vector notation
\begin{eqnarray}
\underline{(A),(B)} \qquad  \qquad
\mbox{div} \; {\bf E} = 0 \; , \qquad \mbox{div} \; {\bf B} = 0 \; ,
\nonumber
\\
\mbox{rot}\; {\bf E} = - \; {\partial \over \partial t}  {\bf B}\;
, \qquad \mbox{rot}\; {\bf B} = +\; {\partial \over \partial t}
{\bf E}\; ,
\label{7.16}
\end{eqnarray}

\noindent where ${\bf E}=(E^{1},E^{2},E^{3}),
{\bf B}=(B^{1},B^{2},B^{3})$.
In the same manner let us  consider   equations
$(A')$ and  $(B')$  (\ref{7.6}) --
 in 3-vector form
 $({\bf A}=(A^{1},A^{2},A^{3})$ they are
\begin{eqnarray}
\underline{(A'), (B') } \qquad \qquad  0 = 0 \; , \qquad {\partial A^{0} \over \partial t}  +
\mbox{div}\; {\bf A} =0 \; , \qquad
\nonumber
\\
  -{\partial {\bf A} \over \partial t}  - \mbox{grad}\; A^{0} = {\bf E} \; , \qquad
\mbox{rot}\; {\bf A} =  {\bf B} \; .
\label{7.18d}
\end{eqnarray}

\section{Massless pseudovector particle , and Lorentz condition}

A pseudovector particle is specified by the relations:
\begin{eqnarray}
\underline{S=\tilde{1}}: \qquad \tilde{\Phi}_{l} \Longrightarrow
\tilde{A}_{l} , \qquad \Phi_{kl} \Longrightarrow F_{kl} \; ,
\nonumber
\\
U(x) = \left [\;  +
           i \; \sigma^{mn}\;  F_{mn}(x) +
           i \; \gamma ^{l} \gamma ^{5} \; \tilde {A}_{l}(x) \; \right ]\; E^{-1} \; ;
\nonumber
\\
(\tilde{A})\qquad i\sigma ^{a} \;  \partial _{a} \; \xi  = m \;   H \; ,
\qquad (\tilde{A}') \qquad i\bar{\sigma}^{a} \; \partial_{a} \;  H = m \;
\xi  \; ,
\nonumber
\\
(\tilde{B})\qquad i \bar{\sigma}^{a} \;  \partial_{a} \; \eta  = m \;
\Delta \; , \qquad (\tilde{B}') \qquad i \sigma ^{a} \; \partial_{a} \;
\Delta     = m \; \eta  \; ;
\nonumber
\\
0=0\; , \qquad
\partial^{l} \tilde{A}_{l} = 0 \; , \qquad
\partial^{l} F_{kl} = 0 \; ,
\nonumber
\\
{1\over 2}\; \epsilon ^{kcmn}\;
\partial _{c} F_{mn}  +
m \;\tilde{A}^{k} = 0 \; , \qquad
 \epsilon ^{dkcl} \;
\partial_{c} \tilde{A}_{l} - m \;F^{dk} = 0 \; .
\label{8.1}
\end{eqnarray}

\noindent
Tensors and spinors are referred by
\begin{eqnarray}
 \left | \begin{array}{cc}
\xi  & \Delta  \\
H & \eta  \end{array} \right | = \left | \begin{array}{cc}
 + \Sigma^{mn} \sigma^{2} \;F_{mn} \;     &
-\; \bar{\sigma}^{l} \sigma^{2} \;  \tilde{A}_{l}  \\[2mm]
-\; \sigma^{l}\sigma^{2} \;   \tilde{A}_{l}   &
 -\; \bar{\Sigma}^{mn} \sigma^{2} \; F_{mn}
\end{array} \right |
\nonumber
\\
 \xi  =
i\sigma^{3} (F_{01} -i \; F_{23}) + (F_{02} -i \; F_{31})
-i\sigma^{1} (F_{03} -i  \;F_{12}) \; ,
\nonumber
\\
 \eta  =
i\sigma^{3} (F_{01} +i \; F_{23}) + (F_{02} +i \; F_{31})
-i\sigma^{1} (F_{03} +i  \;F_{12}) \; ,
\nonumber
\\
\Delta = \left | \begin{array}{rr}
+i\tilde{A}_{1} + \tilde{A}_{2}   &  +i (\tilde{A}_{0} - \tilde{A}_{3} ) \\
-i(\tilde{A}_{0} + \tilde{A}_{3})   & -i\tilde{A}_{1} +
\tilde{A}_{2})
\end{array} \right | ,\;
H = \left | \begin{array}{rr}
-i\tilde{A}_{1} - \tilde{A}_{2}   &  +i (\tilde{A}_{0} + \tilde{A}_{3} ) \\
-i(\tilde{A}_{0} - \tilde{A}_{3})   & +i\tilde{A}_{1} -
\tilde{A}_{2})
\end{array} \right | \; .
\label{8.2
}
\end{eqnarray}

Transition to a massless case is realized as follows:
\begin{eqnarray}
\left. \begin{array}{ll}
(\tilde{A}) & \qquad i\sigma ^{a} \;  \partial _{a} \; \xi  = 0  \\[2mm]
(\tilde{B}) & \qquad i\bar{\sigma}^{a} \; \partial_{a} \; \eta  = 0
\end{array} \right \}
\qquad
 \Longleftrightarrow \qquad
\partial^{l} F_{kl} = 0 \; ,
\qquad {1\over 2}\; \epsilon ^{kcmn}\;
\partial _{c} F_{mn}   = 0  \; ;
\label{8.5}
\\
\left. \begin{array}{ll}
(\tilde{A}') & \qquad i\bar{\sigma} ^{a} \;  \partial _{a} \; H  =  \;  \xi  \\[2mm]
(\tilde{B}') & \qquad i\sigma^{a} \; \partial_{a} \; \Delta  =   \; \eta
\end{array} \right \}
\qquad
 \Longleftrightarrow \qquad
 0=0\; , \qquad
\partial^{l} \tilde{A}_{l} = 0 \; , \qquad
 \epsilon ^{dkcl} \;
\partial_{c} \tilde{A}_{l} = F^{dk} \; .
\label{8.6}
\end{eqnarray}

It should be noted that eqs. (\ref{8.5}) coincide with  (\ref{7.5}), so we need investigate additionally only
equations $(\tilde{A}')$ and $(\tilde{B}')$ -- in vector  form they are
\begin{eqnarray}
\underline{(\tilde{A}'),(\tilde{B}')} \qquad \qquad  0 = 0 \; , \qquad {\partial \tilde{A}^{0} \over \partial t}  +
\mbox{div}\; {\bf \tilde{A}} =0 \; , \qquad
\nonumber
\\
\mbox{rot} \; \vec{\tilde{A}} = + {\bf E} \; , \qquad {\bf B} =
-{\partial {\bf \tilde{A}} \over \partial t}  - \mbox{grad}\;
\tilde{A}^{0}
 \; .
\label{8.12''}
\end{eqnarray}

\section{Comparing results for  vector and pseudovector fields}

Let us collect the results obtained.
Tensor equations for a $S=1$ field are
\begin{eqnarray}
\left. \begin{array}{lll} (\alpha) & \qquad \partial^{l} F_{kl} =
0 \; , & \qquad
\epsilon ^{kcmn}\;\partial _{c} F_{mn}   = 0  \; , \\[2mm]
(\beta)  & \qquad 0=0 \; , \qquad \partial^{a} A_{a} = 0 \; , & \qquad
\partial_{b} A_{c} - \partial_{c} A_{b} = F_{bc}\; ,
\end{array} \right.
\label{9.1a}
\end{eqnarray}
in 3-vector form
\begin{eqnarray}
\left. \begin{array}{lll} (\alpha)  & \qquad \mbox{div} \; {\bf E}
= 0 \; ,
& \qquad \mbox{div} \; {\bf B} = 0 \; , \\[2mm]
& \qquad \mbox{rot}\; {\bf E} = - \; \partial_{t} {\bf B}    \; ,
&
\qquad \mbox{rot}\; {\bf B} = +\; \partial _{t} {\bf E}  \; , \\[2mm]
(\beta)  & \qquad
 0 = 0 \; , & \qquad \partial_{t} A^{0}  + \mbox{div}\; {\bf A} =0 \; , \\[2mm]
 & \qquad  -\partial _{t} {\bf A}  - \mbox{grad}\; A^{0} = \vec{E} \; ,&
 \qquad \mbox{rot}\; {\bf A} = + {\bf B} \; ;
\end{array} \right.
\label{9.1b}
\end{eqnarray}

\noindent
tensor equations for a $S=\tilde{1} $ field
\begin{eqnarray}
\left. \begin{array}{lll} (\tilde{\alpha})  & \qquad
\partial^{l} F_{kl}^{\sim} = 0 \; , & \qquad
\epsilon ^{kcmn}\; \partial _{c} F_{mn}^{\sim}   = 0  \; , \\[2mm]
(\tilde{\beta}) &  \qquad 0=0\; , \qquad \partial^{l}
\tilde{A}_{l} = 0 \; ,& \qquad
 \epsilon _{dk}^{\;\;\;\;\;cl} \; \partial_{c} \tilde{A}_{l} = F_{dk}^{\sim} \; ,
\end{array} \right.
\label{9.2a}
\end{eqnarray}

\noindent  in 3-vector form
\begin{eqnarray}
\left. \begin{array}{lll} (\tilde{\alpha})  &   \qquad
\mbox{div} \; {\bf E} ^{\sim} = 0 \; , & \qquad \mbox{div} \; {\bf B} ^{\sim} = 0 \; ,\\[2mm]
& \qquad \mbox{rot}\; {\bf E} ^{\sim} = - \; \partial _{t} {\bf B} ^{\sim}    \; ,
& \qquad
\mbox{rot}\; {\bf B} ^{\sim} = +\; \partial_{t} {\bf E}  ^{\sim}  \; , \\[2mm]
(\tilde{\beta})  & \qquad 0 = 0 \; , & \qquad \partial_{t}
\tilde{A}^{0}  +
\mbox{div}\; {\bf  \tilde{A}} =0 \; ,\\[2mm]
& \qquad -\partial_{t} {\bf \tilde{A}}  - \mbox{grad}\;
\tilde{A}^{0} =  {\bf B}^{\sim} \; , & \qquad \mbox{rot} \;
{\bf \tilde{A}} = + {\bf E} ^{\sim}   \; .
\end{array} \right.
\label{9.2b}
\end{eqnarray}

\section{Maxwell equations in presence of sources,  fields with different parities}

For the case  $S=1$, spinor massless equations with sources can be obtained from  equation (A),(B) in (\ref{7.1a})
\begin{eqnarray}
S=1 \qquad \left. \begin{array}{ll} (A) & \qquad i\sigma ^{a} \;
\partial _{a} \; \xi (x)  = m \;   H (x) \; ,
 \\[2mm]
(B) & \qquad i\bar{\sigma}^{a} \; \partial_{a} \; \eta (x) = m \;
\Delta (x) \;
\end{array} \right.
\nonumber
\end{eqnarray}

\noindent  by  formal changes
\begin{eqnarray}
m H (x)= - \; i\sigma^{l} \sigma^{2} \; m A_{l} (x)\; ,
 \qquad \Longrightarrow \qquad - j(x) = -i \;\sigma^{k} \sigma^{2} \; j_{k}(x) \; ,
\nonumber
\\
m \Delta (x)  = + \; i\bar{\sigma}^{l} \sigma^{2}\; m A_{l} (x)\;
, \qquad \Longrightarrow \qquad  - \bar{j}(x) = +i
\;\bar{\sigma}^{k} \sigma^{2} \; j_{k}(x) \; .
\nonumber
\end{eqnarray}

\noindent Thus, we get
\begin{eqnarray}
 (S=1) \qquad  \left. \begin{array}{ll}
(A) & \qquad i\sigma ^{a} \;  \partial _{a} \; \xi (x)  = -\; j (x) \; ,  \\[2mm]
(B) & \qquad i\bar{\sigma}^{a} \; \partial_{a} \; \eta (x) =- \;
\bar{j}(x) \; ,
\end{array} \right.
\nonumber
\\
\left. \begin{array}{lll}
    \qquad  & \qquad
\partial^{l} F_{kl} (x) = -\;j_{k} (x) \; , &\qquad
\epsilon^{kcmn} \partial_{c} F_{mn} (x)  = 0 \; ;\\[4mm]
 & \qquad \mbox{div} \; {\bf E} = +j^{0} \; ,
& \qquad \mbox{div} \; {\bf B} = 0 \; , \\[2mm]
& \qquad \mbox{rot}\; {\bf E} = - \; \partial_{t} {\bf B}    \; ,
& \qquad \mbox{rot}\; {\bf B} = +\; \partial _{t} {\bf E}  +
{\bf j} \; ,
\end{array} \right.
\label{10.3a}
\end{eqnarray}

\noindent and remaining equation $(A'),(B')$:
\begin{eqnarray}
 (S=1) \qquad  \left. \begin{array}{ll}
(A') & \qquad i\bar{\sigma} ^{a} \;  \partial _{a} \; H  =  \;  \xi  \\[2mm]
(B') & \qquad i\sigma^{a} \; \partial_{a} \; \Delta  =   \; \eta
\end{array} \right \}
\qquad
 \Longrightarrow
\nonumber
\\
\partial^{l} A_{l} = 0 \; , \qquad \partial_{k} A_{l} - \partial_{l} A_{k} =  \; F_{kl} \; ,
\nonumber
\\
  0 = 0 \; , \qquad {\partial A^{0} \over \partial t}  +
\mbox{div}\; {\bf A} =0 \; , \qquad
\nonumber
\\
  -{\partial {\bf A} \over \partial t}  - \mbox{grad}\; A^{0} = {\bf E} \; , \qquad
\mbox{rot}\; {\bf A} =  {\bf B} \; .
\label{10.3b}
\end{eqnarray}

In the same manner, one should consider the case $S = \tilde{1}$. Making in spinor equations
\begin{eqnarray}
S=\tilde{1}: \qquad \left. \begin{array}{ll} (\tilde{A}) & \qquad i\sigma
^{a} \;  \partial _{a} \; \xi (x)  = m \;   H (x) \; ,
 \\[2mm]
(\tilde{B}) & \qquad i\bar{\sigma}^{a} \; \partial_{a} \; \eta (x) = m \;
\Delta (x)
\end{array} \right.
\nonumber
\end{eqnarray}

\noindent  formal changes
\begin{eqnarray}
m H (x)= - \; \sigma^{k} \sigma^{2} \; m \tilde{A}_{k} (x)\; ,
 \qquad \Longrightarrow \qquad -\; \tilde{j}(x) = - \;\sigma^{k} \sigma^{2} \; \bar{\tilde{j}}_{k}(x) \; ,
\nonumber
\\
m \Delta (x)  = - \; \bar{\sigma}^{k} \sigma^{2}\; m \tilde{A}_{k}
(x)\; , \qquad \Longrightarrow \qquad -\; \bar{\tilde{j}}(x) = -
\;\bar{\sigma}^{k} \sigma^{2} \; \tilde{j}_{k}(x)
\nonumber
\end{eqnarray}

\noindent we arrive at
\begin{eqnarray}
\left. \begin{array}{ll}
(\tilde{A}) & \qquad i\sigma ^{a} \;  \partial _{a} \; \xi (x)  = -\; \tilde{j} (x) \; ,  \\[2mm]
(\tilde{B}) & \qquad i\bar{\sigma}^{a} \; \partial_{a} \; \eta (x) = -\;
\bar{\tilde{j}}(x) \; ,
\end{array} \right.
\nonumber
\\
\left. \begin{array}{lll} (S=\tilde{1}) & \qquad
\partial^{l} F_{kl}^{\sim} = 0 \; , & \qquad
{1\over 2}\; \epsilon ^{kcmn}\; \partial _{c} F_{mn}^{\sim} = + \tilde{j}^{k }  \; , \\[4mm]
&   \qquad
\mbox{div} \; {\bf E}^{\sim} = 0 \; , & \qquad \mbox{div} \; {\bf B}^{\sim} = -\tilde{j}^{0\sim} \; ,\\[2mm]
& \qquad \mbox{rot}\; {\bf E}^{\sim} = - \; \partial_{t}
{\bf B}^{\sim} + \tilde{{\bf  j}} \; , & \qquad \mbox{rot}\;
{\bf B}^{\sim} = +\;
\partial_{t}  {\bf E} ^{\sim} \; .
\end{array} \right.
\label{10.5a}
\end{eqnarray}

\noindent
and remaining equations $(\tilde{A}'),(\tilde{B}')$:
\begin{eqnarray}
(S=\tilde{1}) \qquad \left. \begin{array}{ll}
(\tilde{A}') & \qquad i\bar{\sigma} ^{a} \;  \partial _{a} \; H  =  \;  \xi  \\[2mm]
(\tilde{B}') & \qquad i\sigma^{a} \; \partial_{a} \; \Delta  =   \; \eta
\end{array} \right \}
\qquad
 \Longleftrightarrow
\nonumber
\\
0=0 \; , \qquad
\partial^{l} \tilde{A}_{l} = 0 \; , \qquad
 \epsilon ^{dkcl} \;
\partial_{c} \tilde{A}_{l} = F^{dk\sim} \; ,
\nonumber
\\[3mm]
 0 = 0 \; ,  \qquad \partial_{t}  \tilde{A}^{0}  +
\mbox{div}\; {\bf  \tilde{A}} =0 \; ,
\nonumber
\\
  -\partial_{t} {\bf \tilde{A}}  - \mbox{grad}\;
\tilde{A}^{0} =  {\bf B}^{\sim} \; ,  \qquad \mbox{rot} \;
{\bf \tilde{A}} = + {\bf E}^{\sim}    \; .
\label{10.5b}
\end{eqnarray}

One notices that a simple algebraic summing of the  systems related to $(A),(B)$  and  $(\tilde{A}),(\tilde{B})$ for fields of the types
$S=1$ $S = \tilde{1}$ (see (\ref{10.3a})     and (\ref{10.5a})), one arrives at
equations for Maxwell electromagnetic theory with two charge, electric and magnetic:
\begin{eqnarray}
{\bf E} + {\bf E}^{\sim} = \hat{{\bf E} }\; , \qquad {\bf B} +
{\bf B}^{\sim} = \hat{{\bf B}} \; ,
\nonumber
\\
\mbox{div}\; \hat{{\bf E}} = j^{0} \; , \qquad \mbox{div}\; \hat{{\bf B}} =-
\tilde{j}^{0} \; ,
\nonumber
\\
\mbox{rot}\; \hat{{\bf E} } = -\partial_{t} \hat{{\bf B}} + \tilde{{\bf j}} \; ,
\qquad \mbox{rot}\; \hat{{\bf B}}  = +\partial_{t}  \hat{{\bf E}} + {\bf j} \;
,
\label{10.6a}
\end{eqnarray}

\noindent
4-tensor description of that summing looks
\begin{eqnarray}
( \hat{{\bf E}}, c \hat{{\bf B}}), \qquad \hat{F}_{ab}=  F_{ab} + F_{ab}^{\sim} =
(\partial_{a}A_{b} - \partial_{b} A_{a}) +
\epsilon_{ab}^{\;\;\;\;kl} \partial_{k} \tilde{A}_{l}\;,
\nonumber
\\[2mm]
\partial^{l} \; [\;  F_{kl} (x) +  F_{kl}^{\sim}  \; ] = -\;j_{k} (x) \; ,
\nonumber
\\
{1 \over 2}\; \epsilon^{kcmn} \partial_{c} \; [\; F_{mn} (x)  + F_{mn}^{\sim}  \;] = + \tilde{j}^{k}  \; .
\label{10.6b}
\end{eqnarray}

\noindent which coincides with 2-potential approach to Maxwell electrodynamics with  two charges
\cite{1975-Strazhev-Tom}.

 The above method of obtaining extended Maxwell equations with magnetic charges from two \underline{independent}
Maxwell theories with different intrinsic parities
\begin{eqnarray}
S=1 \; , \qquad   (A, B)\; ,    \qquad (A', B') \; ;
\nonumber
\\
S=\tilde{1} \; , \qquad   (\tilde{A}, \tilde{B}) \; ,  \qquad (\tilde{A}', \tilde{B}' )\; ;
\label{10.7}
\end{eqnarray}

\noindent
may be sketched by the scheme
\begin{eqnarray}
(S=1)   + (S=\tilde{1}):  \qquad  (A,B)  +  (\tilde{A}, \tilde{B})  \; ,\;\;
A', B', \tilde{A}', \tilde{B}' \; .
\label{10.8a}
\end{eqnarray}

\noindent
One could introduce a symmetrical combination by a similar  scheme
\begin{eqnarray}
(S=1)   - (S=\tilde{1}):  \qquad  (A,B)  +  (\tilde{A}, \tilde{B})  \;
A', B', \tilde{A}', \tilde{B}' \; .
\label{10.8b}
\end{eqnarray}

\noindent which corresponds to
\begin{eqnarray}
{\bf E} - {\bf E}^{\sim} = \breve{{\bf E} }\; , \qquad {\bf B} -
{\bf B}^{\sim} = \breve{{\bf B}} \; ,
\nonumber
\\
\mbox{div}\;  \breve{{\bf E}} = j^{0} \; , \qquad \mbox{div}\; \breve{{\bf B}} = +
\tilde{j}^{0} \; ,
\nonumber
\\
\mbox{rot}\; \breve{{\bf E} } = -\partial_{t} \breve{{\bf B}} - \tilde{{\bf j}} \; ,
\qquad \mbox{rot}\; \breve{{\bf B}}  = +\partial_{t}  \breve{{\bf E}} + {\bf j} \;
;
\label{10.9a}
\end{eqnarray}

\noindent
4-tensor description of that summing looks
\begin{eqnarray}
( \breve{{\bf E}}, c \breve{{\bf B}}), \qquad \breve{F}_{ab}=  F_{ab} - F_{ab}^{\sim} =
(\partial_{a}A_{b} - \partial_{b} A_{a}) -
\epsilon_{ab}^{\;\;\;\;kl} \partial_{k} \tilde{A}_{l}\;,
\nonumber
\\[2mm]
\partial^{l} \; [\;  F_{kl} (x) - {F}^{\sim}_{kl}  \; ] = -\;j_{k} (x) \; ,
\nonumber
\\
{1 \over 2}\; \epsilon^{kcmn} \partial_{c} \; [\; F_{mn} (x)  -  F_{mn}^{\sim}  \;] = - \tilde{j}^{k }  \; .
\label{10.6b}
\end{eqnarray}

Evidently, the system consisting of two independent Maxwell models, $S=1$ and $S= \tilde{1}$,
is equivalent  to the pair
\begin{eqnarray}
(S=1)   \pm  (S=\tilde{1}):  \qquad  (A,B)  \pm  (\tilde{A}, \tilde{B})  \; ,\;\;
A', B', \tilde{A}', \tilde{B}' \; .
\label{10.8c}
\end{eqnarray}

Let us call the  doubled model (\ref{10.8c})  an \underline{extended Maxwell theory},
 it includes two photon fields different in parity, and two types of sources, electrical and  magnetic charges.

\section{Extension to the Maxwell theory in Riemannian space-time}

Let us start with a generally covariant tetrad-based Dirac-K\"{a}hler equation in 4-spinor form
\cite{1989-Red'kov}, \cite{2000-Red'kov}:
\begin{eqnarray}
[\;  i \gamma ^{\alpha }(x) \;  ( \partial /\partial x^{\alpha } +
B_{\alpha }(x) ) - m \; ]\; U(x) = 0 \; ,
\label{A.1a}
\end{eqnarray}

\noindent where  $B_{\alpha }(x)$ is a 2-rank bispinor connection
\begin{eqnarray}
B_{\alpha }(x) = {1\over 2} J^{ab} e^{\beta }_{(a)}
 \nabla _{\alpha }(e_{(b)\beta }) =
 \Gamma _{\alpha }(x) \otimes  I + I  \otimes  \Gamma_
{\alpha}(x) \; ,
\nonumber
\end{eqnarray}

\noindent and  $J^{ab} = ( \sigma ^{ab} \otimes  I  + I \otimes
\sigma ^{ab} )$ stands for generators for 2-rank  bispinor under the Lorentz group.
From (\ref{A.1a})  specified in Weyl spinor basis, one derives  the following equations in 2-spinor form:
\begin{eqnarray}
i \sigma ^{\alpha }(x)\; [\; \partial /\partial x^{\alpha } \; +
\; \Sigma _{\alpha }(x) \otimes  I  + I \otimes  \Sigma _{\alpha
}(x)\; ]\;
 \zeta (x) =    m \; H(x) \; ,
\nonumber
\\
i \bar{\sigma}^{\alpha }(x)\; [\; \partial / \partial x^{\alpha }
\; + \;
 \bar{\Sigma}_{\alpha }(x) \otimes  I + I \otimes  \Sigma _{\alpha }(x)\; ] \;  H(x) =
m \; \xi (x) \; ,
\nonumber
\\
i \bar{\sigma}^{\alpha }(x)\; [ \; \partial /\partial x^{\alpha
}\; +\;
 \bar{\Sigma}_{\alpha }(x) \otimes  I +
I \otimes  \bar{\Sigma}_{\alpha }(x) \; ] \; \eta (x) = m \;
\Delta (x) \; ,
\nonumber
\\
i \sigma ^{\alpha }(x)\; [\; \partial /\partial x^{\alpha }\; + \;
\Sigma _{\alpha }(x) \otimes  I +
 I \otimes  \bar{\Sigma}_{\alpha }(x) \;]\; \Delta (x) =
 m \; \eta (x)\; .
\label{A.1b}
\end{eqnarray}

\noindent  Symbols $\Sigma_{\alpha}(x)$ and $\bar{\Sigma}_{\alpha}(x)$
stand for Infeld -- van der Vaerden connections:
\begin{eqnarray}
\sigma ^{\alpha }(x) = \sigma ^{a} \; e^{\alpha }_{(a)}(x) \; ,
\qquad \bar{\sigma }^{\alpha } (x) = \bar{\sigma }^{a} \; e
^{\alpha } _{(a)} (x)\; ,
\nonumber
\\
\Sigma _{\alpha }(x) = {1 \over 2} \; \Sigma ^{ab} \; e^{\beta
}_{(a)} \; \nabla _{\alpha } (e_{(b)\beta }) \;  ,  \qquad
\bar{\Sigma } _{\alpha }(x) = {1 \over 2} \; \bar{\Sigma }^{ab} \;
e^{\beta }_{(x)} \; \nabla _{\alpha } (e_{(b)\beta }) \;   ,
\nonumber
\\
\Sigma ^{a} = {1 \over 4}\; (\;  \bar{\sigma }^{a} \; \sigma ^{b}
\;  - \; \bar{\sigma }^{b} \; \sigma ^{a} \;)\; , \qquad
\bar{\Sigma }^{a}={1  \over  4}  \; ( \; \sigma ^{a}\; \bar{\sigma
}^{b}\; - \;\sigma ^{b} \; \bar{\sigma }^{a}  \; ) \; .
\nonumber
\end{eqnarray}

These spinor equations are equivalent  to  a generally covariant  tensor system
\begin{eqnarray}
\nabla ^{\alpha } \Psi _{\alpha }(x)  + m \Psi (x) = 0 \; , \qquad
\tilde{\nabla }^{\alpha } \Psi _{l}(x) + m \tilde{\Psi} (x)  = 0
\;  ,
\nonumber
\\
\nabla _{\alpha }\Psi (x) + \nabla ^{\beta } \Psi _{\alpha \beta
}(x) -
 m \Psi _{\alpha }(x)= 0 \; ,
\nonumber
\\
 \tilde{\nabla }_{\alpha }\Psi  (x) -
{1\over 2} \epsilon ^{\;\;\beta \rho \sigma }_{\alpha }(x)
 \nabla _{\beta } \Psi _{\rho \sigma }(x) -
m \tilde{\Psi }_{\alpha }(x)  = 0 \; ,
\nonumber
\\
 \nabla _{\alpha } \Psi _{\beta }(x) -
\nabla _{\beta } \Psi _{\alpha }(x) + \epsilon ^{\;\;\;\;\rho
\sigma }_{\alpha \beta }(x) \nabla _{\rho }
 \tilde{\Psi }_{\sigma } (x) - m \Psi _{\alpha \beta }(x) = 0 \; ,
\label{A.1c}
\end{eqnarray}

\noindent where covariant tensor field variables are connected with local tentad tensor variables
by the relations
\begin{eqnarray}
\Psi _{\alpha }(x) = e^{(i)}_{\alpha }(x) \Psi _{i}(x) \; , \qquad
\tilde{\Psi }_{\alpha }(x)= e^{(i)}_{\alpha }(x) \tilde{\Psi
}_{i}(x) \; , \;
\nonumber
\\
\Psi _{\alpha \beta }(x) = e^{(m)}_{\alpha }(x) e^{(n)}_{\beta
}(x) \Psi _{mn}(x)  \; ,
\label{A.2a}
\end{eqnarray}

\noindent and the  Levi-Civita object is determined by
\begin{eqnarray}
\epsilon ^{\alpha \beta \rho \sigma }(x) = \epsilon ^{abcd}
e^{\alpha }_{(a)}(x) e^{\beta }_{(b)}(x)
                 e^{\rho }_{(c)}(x) e^{\sigma }_{(d)}(x)  \; .
\label{4.1.2b}
\end{eqnarray}

\noindent The fields  $\Psi (x) ,\; \Psi _{\alpha }(x) , \; \Psi _{\alpha
\beta }(x)$  are tetrad scalars, and  $\tilde{\Psi }(x), \; \tilde{\Psi }_{\alpha }(x)$  are tetrad pseudoscalars,
and the Levi-Civita object $\epsilon ^{\alpha \beta \rho \sigma }(x)$  is a generally covariant tensor and
a tetrad pseudoscalar.

Now, we should obtain generally covariant equations for ordinary boson of the types
$S=0,\tilde{0}, 1, \tilde{1}$:

\vspace{5mm}

$
S=0, \qquad
\tilde{\Psi }\;  ,  \;\tilde{\Psi }_{\alpha }, \; \Psi _{\alpha \beta } = 0 \; ,
$
\begin{eqnarray}
\qquad
\nabla ^{\alpha } \Psi_{\alpha } + m \Psi  = 0 \; ,
\qquad
\nabla _{\alpha } \Psi  - m \Psi _{\alpha } = 0 \; ,
\qquad
\nabla_{\alpha} \Psi_{\beta} - \nabla_{\beta} \Psi_{\alpha}
= 0 \;  ,
\label{A.3a}
\end{eqnarray}

\noindent
two first  are  the Proca  equations for scalar particle, the last  equation holds identically:
\begin{eqnarray}
(\partial _{\alpha } \; \partial _{\beta }\; \Psi(x) \; - \;
\Gamma ^{\mu }_{\alpha \beta }(x)\; \partial _{\mu } \; \Psi(x) )
\; -  \; ( \partial _{\beta} \; \partial _{\alpha } \; \Psi(x) \;
- \; \Gamma ^{\mu }_{\beta \alpha } (x) \; \partial _{\mu }\;
\Psi(x) ) = 0 \; ,
\nonumber
\end{eqnarray}

\noindent
For a pseudoscalar field we have
\begin{eqnarray}
S= \tilde{0} \; ,\qquad
\nabla ^{\alpha } \tilde{\Psi }_{\alpha }(x) +
 m \tilde{\Psi }(x) = 0 \; ,  \qquad
\nabla _{\alpha } \tilde{\Psi }(x) - m \tilde{\Psi }_{\alpha }(x)
= 0 \; ,
\qquad
  \epsilon_{\alpha \beta}^{\;\;\;\;\rho \sigma} (x) \nabla_{\rho} \Psi_{\sigma}(x) = 0 \; ;
\label{A.3b}
\end{eqnarray}

\noindent here the last equation holds identically.
Now, let  it be  $\Psi _{\alpha }(x) \neq  0 , \; \Psi
_{\alpha \beta }(x) \neq  0$, then
\begin{eqnarray}
S=1\; , \qquad\qquad \nabla^{\alpha} \Psi_{\alpha}(x) = 0 \; , \qquad \nabla ^{\beta }
\Psi _{\alpha \beta }(x) - m \Psi _{\alpha }(x)  =  0 \;  ,
\nonumber
\\
-{1\over 2} \epsilon_{\alpha}^{\;\;\beta \rho \sigma}(x)
\nabla_{\beta} \Psi_{\rho \sigma}(x) = 0 \; ,
\qquad
\nabla_{\alpha} \Psi_{\beta}(x) - \nabla_{\beta}\Psi_{\alpha }(x)
= m \Psi _{\alpha \beta }(x)  \; .
\label{A.4a}
\end{eqnarray}

\noindent Here the first and third equation hold identically:
\begin{eqnarray}
\nabla ^{\alpha } \Psi _{\alpha }(x) = {1\over m} \; \nabla
^{\alpha } \nabla ^{\beta }  \; \Psi _{\alpha \beta }(x) = {1\over
2m}\; [ \; \Psi _{\alpha \nu }(x) \; R^{\nu \;\; \beta \alpha
}_{\;\;\beta }(x) -
\nonumber
\\
- \Psi _{\beta \nu }(x) R^{\nu \;\;  \beta \alpha }_{\;\;\alpha
}(x) \; ] = {1\over 2m} \; [\; - \Psi _{\alpha \nu }(x) \; R^{\nu
\alpha }(x) - \Psi _{\beta \nu }(x) \; R^{\nu \beta }(x)\;  ] = 0 \; ,
\nonumber
\\
-{ 1\over 2m} \; \epsilon ^{\;\;\beta \rho \sigma }_{\alpha }(x)
\;
 \nabla _{\beta } \; [ \; \nabla _{\rho } \Psi _{\sigma }(x) -
                    \nabla _{\sigma } \Psi _{\rho }(x) \; ] =
\nonumber
\\
-{1\over 4m} \; \epsilon ^{\;\;\rho \beta \sigma }_{\alpha }(x) \;
[ \; ( \nabla _{\beta } \nabla _{\rho }(x) -
    \nabla _{\rho } \nabla _{\beta } ) \; \Psi _{\sigma }(x) -
    ( \nabla _{\beta } \nabla _{\sigma }(x) -
    \nabla _{\sigma } \nabla _{\beta } )\;  \Psi _{\rho }(x) \; ] =
\nonumber
\\
= -{1\over 4m} \;  \epsilon ^{\;\;\beta \rho \sigma }_{\alpha
}(x)\;
 ( \Psi ^{\nu } \; R_{\nu \sigma \rho \beta }(x) -
   \Psi ^{\nu }(x) \; R_{\nu \rho \sigma \beta }(x) ) \; = 0 \; .
\end{eqnarray}

\noindent
Now, let  it be  $\Psi (x) =
 \tilde{\Psi }(x) = \Psi _{\alpha }(x) = 0$, then
\begin{eqnarray}
S= \tilde{1}\;, \qquad \qquad \qquad
\nabla^{\alpha} \tilde{\Psi}_{\alpha}(x) = 0 \; , \qquad
\nabla^{\beta} \tilde{\Psi}_{\alpha \beta}(x) = 0 \; ,
\nonumber
\\
{1\over 2}\; \epsilon ^{\;\;\beta \rho \sigma }_{\alpha }(x)\;
\nabla _{\beta } \Psi _{\rho \sigma }(x) + m \;\tilde{\Psi }(x) =
0 \; ,
\qquad
 \epsilon ^{\;\;\;\;\rho \sigma }_{\alpha \beta }(x) \;
\nabla _{\rho } \tilde{\Psi }_{\sigma }(x) - m \;\Psi _{\alpha
\beta }(x) = 0 \; .
\label{A.4b}
\end{eqnarray}

\noindent The first and the second equations  hold identically:
\begin{eqnarray}
\nabla ^{\alpha } \tilde{\Psi }_{\alpha }(x) = -  {1\over 2m}\;
\nabla ^{\alpha } \epsilon ^{\;\;\beta \rho \sigma }_{\alpha }(x)
\; \nabla _{\beta } \; \Psi _{\rho \sigma }(x)  =
 -{1\over 2m}\; \epsilon ^{\;\;\beta \rho \sigma }_{\alpha }(x)
\; \nabla ^{\alpha } \nabla _{\beta } \; \Psi _{\rho \sigma }(x) =
\nonumber
\\
= -{1\over 4m} \epsilon ^{\;\;\beta \rho \sigma }_{\alpha }(x) [\;
\Psi _{\nu \sigma }(x)\;  R^{\nu \;\;\;\; \alpha }_{\;\;\beta \rho
}(x) \; + \; \Psi _{\rho \nu }(x) \; R^{\nu \;\;\;\; \alpha
}_{\;\;\sigma \beta }(x) \; ] \; ,
\nonumber
\\
\nabla ^{\beta } \; \Psi _{\alpha \beta }(x) = {1\over m } \;
\nabla ^{\beta }(x) \; \epsilon ^{\;\;\;\;\rho \sigma }_{\alpha
\beta }(x)\; \nabla _{\rho } \Psi _{\sigma }(x)  =
 {1\over 2m}  \; \epsilon ^{\;\;\rho \sigma }_{\alpha \beta }(x) \;
 \Psi ^{\nu }(x) \; R_{\nu \sigma \rho \beta }(x) \; .
\nonumber
\end{eqnarray}

Constraints separating four boson fields are the same as in the case of Minkowski space:
\begin{eqnarray}
S = 0 \; ,\qquad  \qquad  \tilde{\Delta } = + H\; , \;
\xi  = - \eta \; , \;
      \tilde{\xi } = - \xi\; , \;  \tilde{\eta } = - \eta \; .
\nonumber
\\
S = \tilde{0} \; ,\qquad  \qquad  \tilde{\Delta } = - H
\; , \; \xi  = + \eta \; ,  \;
 \tilde{\xi } = - \xi  \; , \; \tilde{\eta } = - \eta \; .
\nonumber
\\
S = 1 \; , \qquad  \qquad \tilde{\Delta } = + H \; ,
 \; \tilde{\xi } = + \xi \; ,
 \; \tilde{\eta } = + \eta \; .
\nonumber
\\
S = \tilde{1} \; , \qquad \qquad \tilde{\Delta } = - H
\; , \; \tilde{\xi } = + \xi\; ,   \; \tilde{\eta } = + \eta \; .
\label{A.5}
\end{eqnarray}

Without any additional calculation, we can write down spinor and corresponding tensor equations for Maxwell
 theories with  opposite intrinsic parities:

\vspace{5mm}
for a vector model
\begin{eqnarray}
 (S=1) \qquad  \left. \begin{array}{ll}
(A) & \qquad i \sigma ^{\alpha }(x)\; [\; \partial /\partial x^{\alpha } \; +
\; \Sigma _{\alpha }(x) \otimes  I  + I \otimes  \Sigma _{\alpha
}(x)\; ]\;
 \xi (x)  = -\; j (x) \; ,  \\[2mm]
(B) & \qquad  i \bar{\sigma}^{\alpha }(x)\; [ \; \partial /\partial x^{\alpha
}\; +\;
 \bar{\Sigma}_{\alpha }(x) \otimes  I +
I \otimes  \bar{\Sigma}_{\alpha }(x) \; ] \; \eta (x) =- \;
\bar{j}(x) \; ,
\end{array} \right.
\nonumber
\\
\left. \begin{array}{lll}
    \qquad  & \qquad
\nabla^{\beta} F_{\alpha \beta} (x) = -\;j_{\alpha} (x) \; , &\qquad
\epsilon^{\alpha \beta \rho \sigma } \nabla_{\beta} F_{\rho \sigma} (x)  = 0 \; ;
\end{array} \right.
\nonumber
\end{eqnarray}
\begin{eqnarray}
 (S=1) \qquad  \left. \begin{array}{ll}
(A') & \qquad  i \bar{\sigma}^{\alpha }(x)\; [\; \partial / \partial x^{\alpha }
\; + \;
 \bar{\Sigma}_{\alpha }(x) \otimes  I + I \otimes  \Sigma _{\alpha }(x)\; ] \;  H(x)  =  \;  \xi \; , \\[2mm]
(B') & \qquad  i \sigma ^{\alpha }(x)\; [\; \partial /\partial x^{\alpha }\; + \;
\Sigma _{\alpha }(x) \otimes  I +
 I \otimes  \bar{\Sigma}_{\alpha }(x) \;]\; \Delta (x)  =   \; \eta \; ,
\end{array} \right.
\nonumber
\\
\nabla^{\alpha} A_{\alpha} = 0 \; , \qquad \nabla_{\alpha} A_{\beta} - \nabla_{\beta} A_{\alpha} =  \; F_{\alpha \beta} \; ;
\label{A.6}
\end{eqnarray}

\vspace{5mm}

for a pseudovector model
\begin{eqnarray}
S=\tilde{1} \; , \qquad
\left. \begin{array}{ll}
(\tilde{A}) & \qquad i \sigma ^{\alpha }(x)\; [\; \partial /\partial x^{\alpha } \; +
\; \Sigma _{\alpha }(x) \otimes  I  + I \otimes  \Sigma _{\alpha
}(x)\; ]\;
 \xi (x)   = -\; \tilde{j} (x) \; ,  \\[2mm]
(\tilde{B}) & \qquad i \bar{\sigma}^{\alpha }(x)\; [ \; \partial /\partial x^{\alpha
}\; +\;
 \bar{\Sigma}_{\alpha }(x) \otimes  I +
I \otimes  \bar{\Sigma}_{\alpha }(x) \; ] \; \eta (x) = -\;
\bar{\tilde{j}}(x) \; ,
\end{array} \right.
\nonumber
\\
\left. \begin{array}{lll} (S=\tilde{1}) & \qquad
\nabla^{\beta} F_{\alpha\beta}^{\sim} = 0 \; , & \qquad
{1\over 2}\; \epsilon ^{\alpha \beta \rho \sigma }\; \nabla _{\beta} F_{\rho \sigma}^{\sim} = + \tilde{j}^{\alpha}  \; ;
\end{array} \right.
\nonumber
\end{eqnarray}
\begin{eqnarray}
(S=\tilde{1}) \; , \qquad \left. \begin{array}{ll}
(\tilde{A}') & \qquad i \bar{\sigma}^{\alpha }(x)\; [\; \partial / \partial x^{\alpha }
\; + \;
 \bar{\Sigma}_{\alpha }(x) \otimes  I + I \otimes  \Sigma _{\alpha }(x)\; ] \;  H(x)   =  \;  \xi  \\[2mm]
(\tilde{B}') & \qquad i \sigma ^{\alpha }(x)\; [\; \partial /\partial x^{\alpha }\; + \;
\Sigma _{\alpha }(x) \otimes  I +
 I \otimes  \bar{\Sigma}_{\alpha }(x) \;]\; \Delta (x)  =   \; \eta
\end{array} \right.
\nonumber
\\
0=0 \; , \qquad
\nabla^{\alpha} \tilde{A}_{\alpha} = 0 \; , \qquad
 \epsilon ^{\alpha \beta \rho \sigma } \;
\nabla_{\rho} \tilde{A}_{\sigma} = F^{\alpha \beta\sim}  \; .
\label{A.7}
\end{eqnarray}

\section{Symmetry in extended Maxwell theory, duality transformation}

Summing and subtracting respective tensor equations in (\ref{A.6}) and (\ref{A.7}) we arrive at
\begin{eqnarray}
\nabla^{\beta}  [\; F_{\alpha \beta} (x) +   \tilde{F}_{\alpha\beta} (x)\; ] = -\;j_{\alpha} (x) \; ,
\qquad \qquad \qquad
\nabla^{\beta}  [\; F_{\alpha \beta} (x)  -   \tilde{F}_{\alpha\beta} (x)\; ] = -\;j_{\alpha} (x) \; ,
\nonumber
\\
{1\over 2}\; \epsilon ^{\alpha \beta \rho \sigma } (x)\; \nabla _{\beta} \;
[\; F_{\rho \sigma} +   \tilde{F}_{\rho \sigma} \; ]=  + \tilde{j}^{\alpha } (x)\; , \qquad
{1\over 2}\; \epsilon ^{\alpha \beta \rho \sigma } (x)\; \nabla _{\beta} \;
[\; F_{\rho \sigma} -  \tilde{F}_{\rho \sigma} \; ]= - \tilde{j}^{\alpha } (x)\; ,
\nonumber
\\
 F_{\alpha \beta} (x) \pm  \tilde{F}_{\alpha \beta}(x)  =\nabla_{\alpha} A_{\beta}(x) - \nabla_{\beta} A_{\alpha}(x)  \pm
 \epsilon _{\alpha \beta}^{\;\;\;\; \rho \sigma } (x)\; \nabla_{\rho} \tilde{A}_{\sigma} (x) \; ,
\nonumber
\\
\nabla^{\alpha} A_{\alpha} (x)= 0 \; , \qquad \nabla^{\alpha} \tilde{A}_{\alpha}(x) = 0 \; .
\label{A.8}
\end{eqnarray}

\noindent
Here, $F_{\alpha \beta}(x), A_{\alpha}(x), j_{\alpha}(x)$ are tetrad scalar, and
$\epsilon_{\alpha \beta \rho \sigma}(x), \tilde{A}_{\alpha}, \tilde{j}_{\alpha}$
are  tetrad pseudoscalars.

Let us introduce \underline{dual variables} (marked by  a star symbol):
\begin{eqnarray}
F^{*}_{\alpha \beta}(x) =  {1 \over 2} \;
\epsilon_{\alpha \beta }^{\;\;\;\;\rho \sigma}(x)  \; F_{\rho \sigma}(x)\; , \qquad
F_{\alpha \beta}(x) =  -{1 \over 2} \;
\epsilon_{\alpha \beta }^{\;\;\;\;\mu \nu}(x) \;
F^{*}_{\mu \nu}(x)\; ,
\nonumber
\\
\tilde{F}^{*}_{\alpha \beta}(x) =  {1 \over 2} \;
\epsilon_{\alpha \beta }^{\;\;\;\;\rho \sigma}(x)  \; \tilde{F}_{\rho \sigma}(x)\; , \qquad
\tilde{F}_{\alpha \beta}(x) =  -{1 \over 2} \;
\epsilon_{\alpha \beta }^{\;\;\;\;\mu \nu}(x) \;
\tilde{F}^{*}_{\mu \nu}(x)\; .
\label{A.9}
\end{eqnarray}

\noindent One can easily prove an identity
\begin{eqnarray}
 {1 \over 2}\; \epsilon_{\rho \sigma}^{\;\;\;\;\alpha \beta} \; [F_{\alpha \beta} (x) \pm  \tilde{F}^{\alpha \beta}(x) ] =
 {1 \over 2}\; \epsilon_{\rho \sigma}^{\;\;\;\;\alpha \beta} \;  [ \nabla_{\alpha} A_{\beta}(x) - \nabla_{\beta} A_{\alpha}(x)  \pm
 \epsilon _{\alpha \beta}^{\;\;\;\; \delta \gamma } (x)\; \nabla_{\delta} \tilde{A}_{\gamma} (x) \;] =
 \nonumber
 \\
 = \epsilon_{\rho \sigma}^{\;\;\;\;\alpha \beta} \nabla_ {\alpha} A_{\beta}  \mp (\nabla_{\rho} \tilde{A}_{\sigma} -
 \nabla_{\sigma}  \tilde{A}_{\rho}  )\; ,
 \nonumber
 \end{eqnarray}

\noindent that is
\begin{eqnarray}
F_{\rho \sigma}^{*} \pm  \tilde{F}_{\rho \sigma}^{*} = \mp \;
 (\nabla_{\rho} \tilde{A}_{\sigma} -
 \nabla_{\sigma}  \tilde{A}_{\rho}  ) +
 \epsilon_{\rho \sigma}^{\;\;\;\;\alpha \beta} \nabla_ {\alpha} A_{\beta} \;  \; .
\label{A.10}
\end{eqnarray}

\noindent
Thus, the system of extended Maxwell theory may be  presented as follows:
\begin{eqnarray}
\nabla^{\beta}  [\; F_{\alpha \beta} (x) +   \tilde{F}_{\alpha\beta} (x)\; ] = -\;j_{\alpha} (x) \; ,
\qquad \qquad
\nabla^{\beta}  [\; F_{\alpha \beta} (x)  -   \tilde{F}_{\alpha\beta} (x)\; ] = -\;j_{\alpha} (x) \; ,
\nonumber
\\
 \nabla ^{\beta} \;
[\; F^{*}_{\alpha \beta} +   \tilde{F}^{*}_{\alpha \beta} \; ]=  + \tilde{j}_{\alpha } (x)\; , \qquad
\qquad \qquad
 \nabla _{\beta} \;
[\; F^{*\alpha \beta} -   \tilde{F}^{*\alpha \beta} \; ]= - \tilde{j}_{\alpha } (x)\; ,
\nonumber
\\
 F_{\alpha \beta} (x) \pm  \tilde{F}_{\alpha \beta}(x)  =\nabla_{\alpha} A_{\beta}(x) - \nabla_{\beta} A_{\alpha}(x)  \pm
 \epsilon _{\alpha \beta}^{\;\;\;\; \rho \sigma } (x)\; \nabla_{\rho} \tilde{A}_{\sigma} (x) \; ,
\nonumber
\\
F_{\rho \sigma}^{*} (x)  \pm  \tilde{F}_{\rho \sigma}^{*}  (x) = \mp \;
 (\nabla_{\rho} \tilde{A}_{\sigma}(x)  -
 \nabla_{\sigma}  \tilde{A}_{\rho}(x)  ) +
 \epsilon_{\rho \sigma}^{\;\;\;\;\alpha \beta} (x)  \nabla_ {\alpha} A_{\beta} (x)\;  .
\nonumber
\\
\nabla^{\alpha} A_{\alpha} (x)= 0 \; , \qquad \nabla^{\alpha} \tilde{A}_{\alpha}(x) = 0 \; .
\label{A.11}
\end{eqnarray}

\noindent With the  notation
\begin{eqnarray}
F_{\alpha \beta} (x) +   \tilde{F}_{\alpha\beta} (x) = F^{+}_{\alpha \beta}(x) \; ,\qquad
F^{*}_{\alpha \beta} (x) +   \tilde{F}^{*}_{\alpha\beta} (x) = F^{+*}_{\alpha \beta}(x) \; ,
\nonumber
\\
 F_{\alpha \beta} (x) -   \tilde{F}_{\alpha\beta} (x) = F^{-}_{\alpha \beta}(x) \; , \qquad
 F^{*}_{\alpha \beta} (x) -   \tilde{F}^{*}_{\alpha\beta} (x) = F^{-*}_{\alpha \beta}(x) \; ,\nonumber
\label{A.12}
\end{eqnarray}

\noindent eqs. (\ref{A.11}) look
\begin{eqnarray}
\nabla^{\beta}   F^{+}_{\alpha \beta} (x)  = -\;j_{\alpha} (x) \; ,
\qquad \qquad
\nabla^{\beta}   F^{-}_{\alpha \beta} (x)  = -\;j_{\alpha} (x) \; ,
\nonumber
\\
 \nabla ^{\beta} \;
F^{+*}_{\alpha \beta} =  + \tilde{j}^{\alpha } (x)\; , \qquad
\qquad \qquad
 \nabla ^{\beta} \;
 F^{-*}_{\alpha \beta} = - \tilde{j}^{\alpha } (x)\; ,
\nonumber
\\
 F^{+}_{\alpha \beta} (x)  =\nabla_{\alpha} A_{\beta}(x) - \nabla_{\beta} A_{\alpha}(x)  +
 \epsilon _{\alpha \beta}^{\;\;\;\; \rho \sigma } (x)\; \nabla_{\rho} \tilde{A}_{\sigma} (x) \; ,
\nonumber
\\
 F^{-}_{\alpha \beta} (x)  =\nabla_{\alpha} A_{\beta}(x) - \nabla_{\beta} A_{\alpha}(x)  -
 \epsilon _{\alpha \beta}^{\;\;\;\; \rho \sigma } (x)\; \nabla_{\rho} \tilde{A}_{\sigma} (x) \; ,
\nonumber
\\
F^{+*}_{\rho \sigma} (x)   = - \;
 (\nabla_{\rho} \tilde{A}_{\sigma}(x)  -
 \nabla_{\sigma}  \tilde{A}_{\rho}(x)  ) +
 \epsilon_{\rho \sigma}^{\;\;\;\;\alpha \beta} (x)  \nabla_ {\alpha} A_{\beta} (x)\;  .
\nonumber
\\
F^{-*}_{\rho \sigma} (x)   = + \;
 (\nabla_{\rho} \tilde{A}_{\sigma}(x)  -
 \nabla_{\sigma}  \tilde{A}_{\rho}(x)  ) +
 \epsilon_{\rho \sigma}^{\;\;\;\;\alpha \beta} (x)  \nabla_ {\alpha} A_{\beta} (x)\;  .
\nonumber
\\
\nabla^{\alpha} A_{\alpha} (x)= 0 \; , \qquad \nabla^{\alpha} \tilde{A}_{\alpha}(x) = 0 \; .
\label{A.13}
\end{eqnarray}

The system  obtained is invariant under the following linear transformation (extended duality operation)
\begin{eqnarray}
A_{\alpha}  \Longrightarrow  - \tilde{A}'_{\alpha} \; , \qquad
\tilde{A}_{\alpha} \Longrightarrow  + A'_{\alpha}\; ,
\nonumber
\\
F^{+}_{\alpha \beta}   \Longrightarrow  + F^{'+*}_{\alpha \beta} \; , \qquad
F^{+*}_{\alpha \beta}   \Longrightarrow  - F^{'+}_{\alpha \beta} \; ,
\nonumber
\\
F^{-}_{\alpha \beta}   \Longrightarrow  - F^{'-*}_{\alpha \beta} \; , \qquad
F^{-*}_{\alpha \beta}   \Longrightarrow  + F^{'-}_{\alpha \beta} \; ,
\nonumber
\\
j_{\alpha} \Longrightarrow - \tilde{j}'_{\alpha} \; , \qquad
\tilde{j}_{\alpha} \Longrightarrow + j'_{\alpha} \; .
\label{A.14}
 \end{eqnarray}

Also, one can note another invariance transformation:
\begin{eqnarray}
A_{\alpha}  \Longrightarrow  + \tilde{A}'_{\alpha} \; , \qquad
\tilde{A}_{\alpha} \Longrightarrow  - A'_{\alpha}\; ,
\nonumber
\\
F^{+}_{\alpha \beta}   \Longrightarrow  - F^{'+*}_{\alpha \beta} \; , \qquad
F^{+*}_{\alpha \beta}   \Longrightarrow  + F^{'+}_{\alpha \beta} \; ,
\nonumber
\\
F^{-}_{\alpha \beta}   \Longrightarrow  + F^{'-*}_{\alpha \beta} \; , \qquad
F^{-*}_{\alpha \beta}   \Longrightarrow  - F^{'-}_{\alpha \beta} \; ,
\nonumber
\\
j_{\alpha} \Longrightarrow + \tilde{j}'_{\alpha} \; , \qquad
\tilde{j}_{\alpha} \Longrightarrow - j'_{\alpha} \; .
\label{A.15}
 \end{eqnarray}

Relations  (\ref{A.14}) and (\ref{A.15})  are particular cases of the continuous extended dual transformation over
4-vector and 4-pseudovector:
\begin{eqnarray}
\cos \chi A_{\alpha} + \sin \chi \tilde{A}_{\alpha} = A'_{\alpha} \; , \qquad
\cos \chi A'_{\alpha} - \sin \chi \tilde{A}'_{\alpha} = A_{\alpha} \; ,
\nonumber
\\
-\sin \chi A_{\alpha} + \cos \chi \tilde{A}_{\alpha} = \tilde{A}'_{\alpha} \; , \qquad
\sin \chi A'_{\alpha} + \cos \chi \tilde{A}'_{\alpha} = \tilde{A}_{\alpha} \; ,
\label{A.16}
\end{eqnarray}

\noindent which generates the following transformation over strength tensors:
\begin{eqnarray}
 F^{+}_{\alpha \beta} (x)  = \nabla_{\alpha} A_{\beta}(x) - \nabla_{\beta} A_{\alpha}(x)  +
 \epsilon _{\alpha \beta}^{\;\;\;\; \rho \sigma } (x)\; \nabla_{\rho} \tilde{A}_{\sigma} (x)=
 \nonumber
 \\
 =
\nabla_{\alpha} (\cos \chi A'_{\beta} - \sin \chi \tilde{A}'_{\beta}) - \nabla_{\beta}
(\cos \chi A'_{\alpha} - \sin \chi \tilde{A}'_{\alpha})  +
 \epsilon _{\alpha \beta}^{\;\;\;\; \rho \sigma } (x)\; \nabla_{\rho} (\sin \chi A'_{\sigma} +
 \cos \chi \tilde{A}'_{\sigma} ) (x) \; ,
 \nonumber
 \end{eqnarray}

 \noindent that is
\begin{eqnarray}
F^{+}_{\alpha \beta}
 = \cos \chi \;  F^{'+}_{\alpha \beta}  + \sin \chi \;  F^{'+*}_{\alpha \beta}  \; ;
 \label{A.17}
\end{eqnarray}

\noindent
analogously
\begin{eqnarray}
F^{+*}_{\alpha \beta}  =  -\sin \chi \; F^{'+}_{\alpha \beta}  + \cos \chi  \; F^{'+*}_{\alpha \beta}  \; .
\label{A.18}
\ \end{eqnarray}

\noindent
In the same manner one derives
\begin{eqnarray}
 F^{-}_{\alpha \beta}   =\cos \chi \;  F^{'-}_{\alpha \beta}  - \sin \chi \;  F^{'-*}_{\alpha \beta}  \; ,
 \qquad F^{-*}_{\alpha \beta}   = \sin \chi  \; F^{'-}_{\alpha \beta}  + \cos \chi  \;  F^{'-*}_{\alpha \beta}  \; .
 \label{A.19}
 \end{eqnarray}

\noindent
Continuous dual transformations over currents are
\begin{eqnarray}
\cos \chi j_{\alpha} + \sin \chi \tilde{j}_{\alpha} = j'_{\alpha} \; , \qquad
\cos \chi j'_{\alpha} - \sin \chi \tilde{j}'_{\alpha} = j_{\alpha} \; ,
\nonumber
\\
-\sin \chi j_{\alpha} + \cos \chi \tilde{j}_{\alpha} = \tilde{j}'_{\alpha} \; , \qquad
\sin \chi j'_{\alpha} + \cos \chi \tilde{j}'_{\alpha} = \tilde{j}_{\alpha} \; .
\label{A.19}
\end{eqnarray}

\section{Acknowledgement}

Let us summarize results.
From the 16-component Dirac-K\"{a}hler field theory, spinor equations  for two types of
massless  vector  photon fields with different parities have been  derived. Their equivalent tensor
equations in terms of the strength tensor $F_{ab}$ and respective 4-vector  $A_{b}$ and 4-pseudovector  $\tilde{A}_{b}$
depending   on intrinsic photon parity are derived; they include additional   sources, electric 4-vector
$j_{b}$ and magnetic 4-pseudovector $\tilde{j}_{b}$. The theories of two types of photon fields
are explicitly uncoupled, their linear combination through summing or subtracting results in
Maxwell electrodynamics  with electric and magnetic charges in 2-potential approach.
The whole analysis is extended  straightforwardly to a curved space-time background.

This  work was  supported  by Fund for Basic Research of Belarus
and JINR  F06D-006.

\end{document}